\documentclass[twocolumn,prl,superscriptaddress,showpacs,preprintnumbers,amsmath,amssymb]{revtex4-1}
\usepackage{graphicx}
\usepackage{color}

\begin{document}

\title{Spontaneous Symmetry Breaking in 2D: Kibble-Zurek Mechanism in Temperature Quenched Colloidal Monolayers}

\author{Patrick Dillmann}
\affiliation{University of Konstanz, D-78457 Konstanz, Germany}
\author{Georg Maret}
\affiliation{University of Konstanz, D-78457 Konstanz, Germany}
\author{Peter Keim}
\email{peter.keim@uni-konstanz.de\\}
\affiliation{University of Konstanz, D-78457 Konstanz, Germany}

\date{\today}

\begin{abstract}
The Kibble-Zurek mechanism describes the formation of topological defects during spontaneous symmetry breaking for quite different systems. Shortly after the big bang, the isotropy of the Higgs-field is broken during the expansion and cooling of the universe. Kibble proposed the formation of monopoles, strings, and membranes in the Higgs field since the phase of the symmetry broken field can not switch globally to gain the same value everywhere in space. Zurek pointed out that the same mechanism is relevant for second order phase transitions in condensed matter systems. Every finite cooling rate induces the system to fall out of equilibrium which is due to the critical slowing down of order parameter fluctuations: the correlation time diverges and the symmetry of the system can not change globally but incorporates defects between different domains. Depending on the cooling rate the heterogeneous order parameter pattern are a fingerprint of critical fluctuations. In the present manuscript we show that a monolayer of superparamagnetic colloidal particles is ideally suited to investigate such phenomena. In thermal equilibrium the system undergos continuous phase transitions according KTHNY-theory. If cooled rapidly across the melting temperature the final state is a polycrystal. We show, that the observations can not be explained with nucleation of a supercooled fluid but is compatible with the Kibble-Zurek mechanism.
\end{abstract}

\pacs{05.70.Fh, 05.70.Ln, 64.60.Q-, 64.70.pv, 82.70.Dd}

\maketitle


The Kibble-Zurek mechanism is a beautiful example, how a hypothesis is strengthened if the underlying concept is applicable at completely different scales. Based on spontaneous symmetry breaking, which is relevant at cosmic \cite{Kibble1980}, atomistic (including suprafluids) \cite{Landau1937,Ginzburg1950,Zurek1985}, and elementary particle scale \cite{Nambu1961,Higgs1964}, the Kibble-Zurek mechanism describes domain and defect formation in the symmetry broken phase.

In big bang theory, the isotropy of a n-component scalar Higgs-field $\Phi_n$ is broken during the expansion and cooling of the universe. Regions in space which are separated by causality can not gain the same phase of the symmetry broken field a priory. The formation of membranes, separating the Higgs fields with different phases \cite{Zeldovich1974}, strings \cite{Vilenkin1981,Albrecht1985} and monopoles \cite{Hooft1974,Polyakov1974} is the consequence \cite{Kibble1976,Kibble1980}. Inflation \cite{Guth1981,Linde1982,Albrecht1982}, recently supported by b-mode polarization of the cosmic background ration (the BICEP2 experiment) \cite{BICEP2}, explains why such defects are so dilute, that they are not observed within the visible horizon of the universe. A frequently used (but not very precise) analogy is the formation of domains separated by grain boundaries in a ferromagnet below the Curie-temperature. The analogy is not precise since the ground state (or true vacuum) of the Higgs field would be a mono-domain whereas a ferromagnet has to be poly-domain to reduce the macroscopic magnetic energy. The standard Higgs field order parameter expansion is visualized with the 'Mexican-Sombrero' such being compatible with second order phase transition. To slow down the transition (to dilute monopoles and to solve the flatness-problem in cosmology \cite{Dicke1979}) a dip in the central peak (false vacuum) was postulated \cite{Kibble1980,Guth1981} in the first models of inflation. Due to the dip the 'supercooled' Higgs-field is separated with a finite barrier from the ground state, compatible with a first order scenario. Hence defects easily arise during the nucleation of grains of differently oriented phases being separated by causality.

Zurek pointed out that spontaneous symmetry braking mediated by critical fluctuations of second order phase transitions will equally lead to topological defects in condensed matter physics \cite{Zurek1985}. While a scalar order parameter (or one component Higgs field $\Phi_1$) can only serve grain boundaries as defects (compare e.g. an Ising model), two component (or complex) order parameters $\Phi_2$ offers the possibility to investigate membranes, strings and monopoles as topological defects depending on the homotopy groups \cite{Kibble1980,Chuang1991}. Examples are liquid crystals and superfluid $He^4$ and Zurek discussed finite cooling rates \cite{Zurek1993} as well as instantaneous quenches \cite{Zurek1996}. For the latter the system has to fall out of equilibrium due to the critical slowing down of order parameter fluctuations close to the transition temperature which is true for every finite but nonzero cooling rate. The symmetry of the order parameter can not change globally since the correlation time diverges. The role of the causality is incurred by the maximum velocity for separated regions to communicate. In condensed matter systems this is the sound velocity (or second sound in suprafluid $He$ defining a 'sonic horizon'. The order parameter pattern at the so called 'fall out time' serves as a fingerprint of the frozen out critical fluctuations with well defined correlation length. After the fall out time, the broken symmetry phase will grow on expense of the residual high symmetry phase. The domain size at the fall out time is predicted to grow algebraically as function of inverse cooling rate \cite{Zurek1993}.

Experiments have been done in $He^4$ for the Lambda-transition, but the detection of defects (a network of vortex lines with quantized rotation) can only be done indirectly \cite{Hendry1994}. Nematic liquid crystals offer the possibility to visualize defects directly with cross polarization microscopy \cite{Chuang1991,Bowick1994} but the underling phase transition is first order \cite{Onsager1949}. In two dimensions, quenched dusty plasma have been investigated experimentally \cite{Hartmann2010,Su2012,Yang2012} and colloidal systems with diameter-tunable microgel spheres \cite{Wang2010} but were not interpreted with respect to Kibble-Zurek mechanism. A careful theoretical study of a quenched 2D XY-model is given in \cite{Jelic2011}.

In the present manuscript we show that a monolayer of super-paramagnetic particles is ideally suited to investigate the Kibble-Zurek mechanism in two dimensions for structural phase transitions. In thermal equilibrium the system undergos two transitions with an intermediate hexatic phase according to the KTHNY-theory \cite{Kosterlitz1972,Kosterlitz1973,Halperin1978,Nelson1979,Young1979}. Thus, this theory predicts two transition temperatures (for orientational and translational symmetry breaking) and calculates critical exponents for both diverging correlation lengths. The divergences are exponential (a unique feature of 2D systems) as predicted by renormalization group theory \cite{Halperin1978,Young1979} and determined in experiment \cite{Keim2007}. The exponential divergence (one can take the derivatives arbitrarily often) and the marginal "hump" in the specific heat \cite{Deutschlaender2014} is the reason for naming this transition continuous (and not second order).
\begin{figure}
  \includegraphics[width=0.9\linewidth]{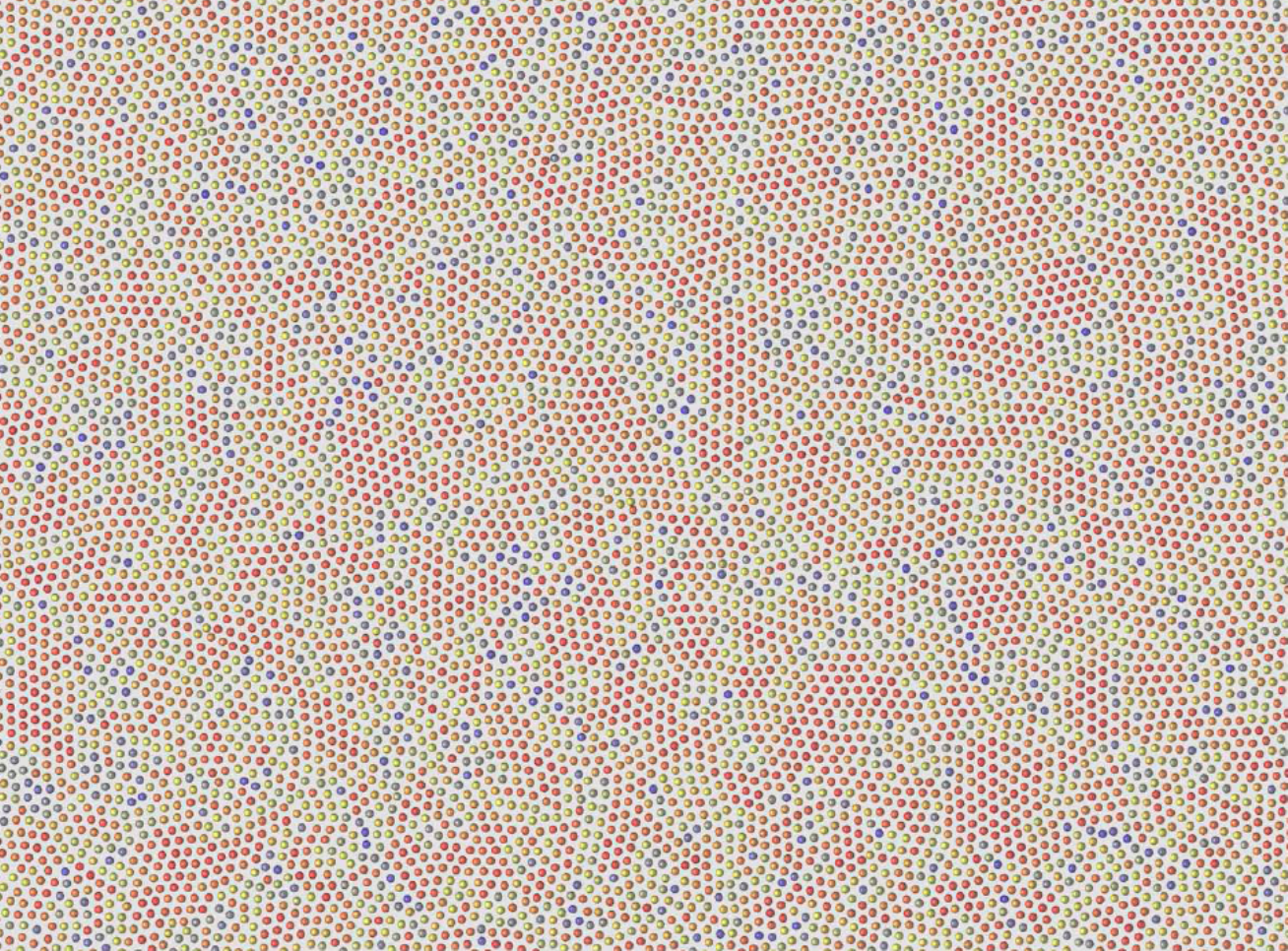}
 \caption{\label{fig01} Snapshot of the monolayer with symmetry broken domains, $120\;\mbox{s}$ after a quench from $\Gamma=14$ to a final coupling strength of $\Gamma_E=140$. The color code (from blue $m=0$ over yellow to red $m=1$) is given by the magnitude $m$ of the local bond order parameter. Blue particles are disordered (low order = high symmetry phase) and red particles have large sixfold symmetry indicating domains with broken symmetry.
 }
\end{figure}

The 2D colloidal monolayer is realized out as follows. A droplet of a suspension composed of polystyrene spheres dispersed in water is suspended by surface tension in a top sealed cylindrical hole ($\varnothing = 6\;mm$) of a glass plate. The particles are $4.5\;\mu m$ in diameter and have a mass density of $1.5\;g/cm^3$ leading to sedimentation. The large density is due to the fact that the polystyrene beads are doped with iron oxide nano-particles leading to super-paramagnetic behavior of the colloids. After sedimentation particles are arranged in a monolayer at the planar and horizontal water-air interface of the droplet and hence form an ideal 2D system. Particles are small enough to perform Brownian motion but large enough to be monitored with video-microscopy. An external magnetic field $H$ perpendicular to the water-air interface induces a magnetic moment in each bead (parallel to the applied field) leading to repulsive dipole-dipole interaction between all particles. We use the dimensionless control parameter $\Gamma$ to characterize this interaction strength. $\Gamma$ is given by the ratio of dipolar magnetic energy to thermal energy
\begin{equation}
\Gamma=\frac{\mu_0}{4\pi}\frac{(\chi H)^2(\pi \rho)^{3/2}}{k_BT} \propto T_{sys}^{-1}
\end{equation}
and thus can be regarded as an inverse system temperature. The state of the system in thermal equilibrium - liquid, hexatic, or solid - is solely defined by the strength of the magnetic field $H$ since the temperature $T$, the 2D particle density $\rho$ and the magnetic susceptibility per bead $\chi$ are kept constant experimentally. In these units the inverse melting temperature (crystalline - hexatic) is at about $\Gamma_m = 60$ and the transition from hexatic to isotropic at about $\Gamma_i = 57$ \cite{commentonTm}. Since the system temperature is given by an outer field, enormous cooling rates are accessible compared to atomic systems. Based on a well equilibrated liquid system \cite{equilibration_time} at $\Gamma\approx15$ we initiate a temperature jump with cooling rates up to $d\Gamma/dt \approx10^4~s^{-1}$ into the crystalline region of the phase diagram $\Gamma_M\geq60$. This temperature quench triggers the solidification within the whole monolayer (the heat transfer is NOT done through the surface of the material as usually in 3D condensed matter systems) and time scale of cooling is $10^5$ faster compared to fastest intrinsic scales, e.g. the Brownian time $\tau_B = 50\;\mbox{sec}$, particles need to diffuse the distance of their own diameter. An elaborate description of the experimental setup can be found in \cite{Ebert2009}. The field of view is $1160 \times 865\;\mu m^2$ in size and contains about $9000$ colloids, the whole monolayer contains of $\sim 300000$ particles. Each temperature quench is repeated at least ten times to the same final value of the control parameter $\Gamma_F$ with sufficient equilibration times in between. Fig \ref{fig01} shows the monolayer two minutes after the quench from deep in the fluid phase below the melting temperature (above the control parameter $\Gamma_m$), (see also movie bondordermagnitude.avi of the supplemental material). The color code of particle $l$ at position $\vec{r}_l$ is given by the magnitude $ m_l = m(\vec{r_l}) = \psi_l^\ast\psi_l$ of the local complex (six-folded) bond order field $\psi_l = 1/ N_j \sum_{k=1}^{N_j} e^{i 6\theta_{kl}}$ given by the $N_j$ nearest neighbors. Blue particles have low bond orientational order $m \approx 0$ (high isotropic symmetry) and red particles indicate domains of broken symmetry $m \approx 1$ (high local sixfold order). Note that we investigate the orientational order and not the translational one, since a) we can hardly distinguish between 'poly-hexalline' and poly-crystalline on finite length scales, b) the hexatic phase is narrow compared to the quench depth, and c) translational order parameters are based on reciprocal lattice vectors $\vec{G}$ being ill defined in poly-crystalline samples. The hexatic phase is not observable as function of time after a quench \cite{Dillmann2008} and the bond-order correlation function
\begin{equation}
\label{g6} g_6(r) = \langle |\psi(\vec{r_k})\psi^\ast(\vec{r_j})|
\rangle_{kj} = \langle |\psi(\vec{r})\psi^\ast(\vec{0})| \rangle
\quad ,
\end{equation}
always decays exponential $g_6(r,t) \sim \exp(r/\xi_6(t)) $. The orientational correlation length $\xi_6(t)$ grows monotonically after the quench and the final state is poly-crystalline. As a side remark; the Kibble-Zurek mechanism implies that poly-crystallinity can not solely be used as argument for first order phase transitions since zero cooling rates can not be realized during preparation.
\begin{figure}
\includegraphics[width=1.\linewidth]{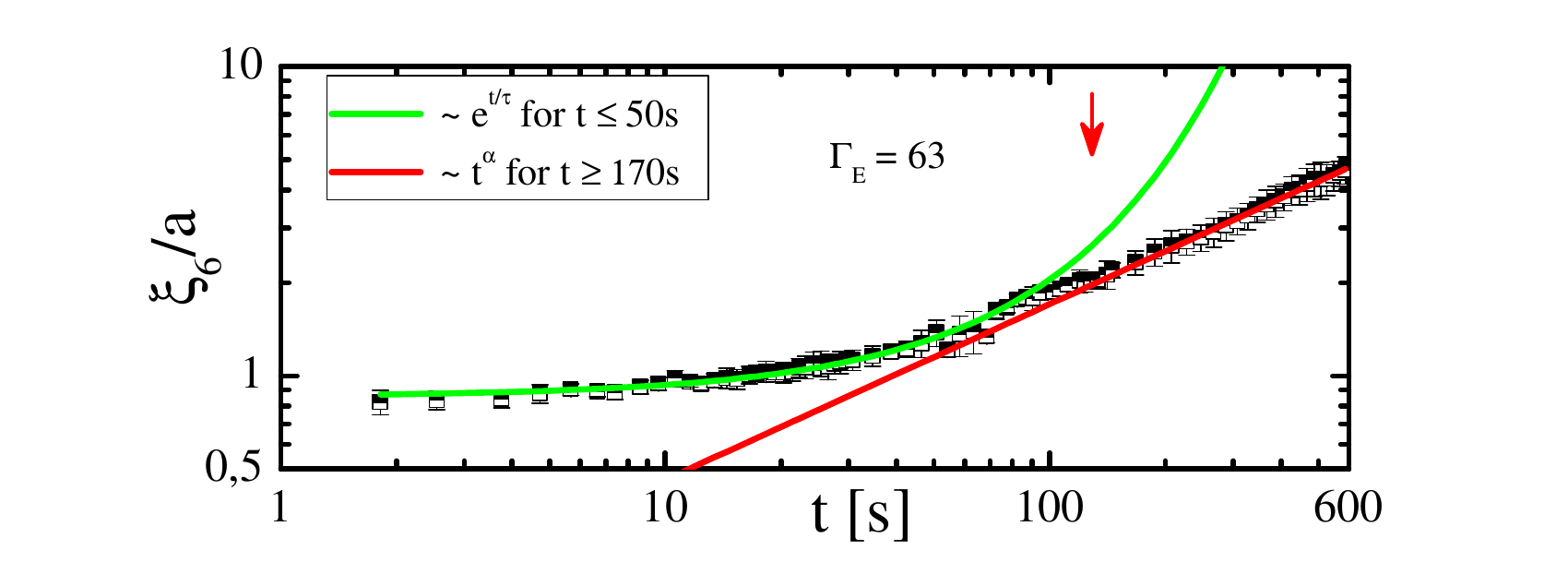}
\includegraphics[width=1.\linewidth]{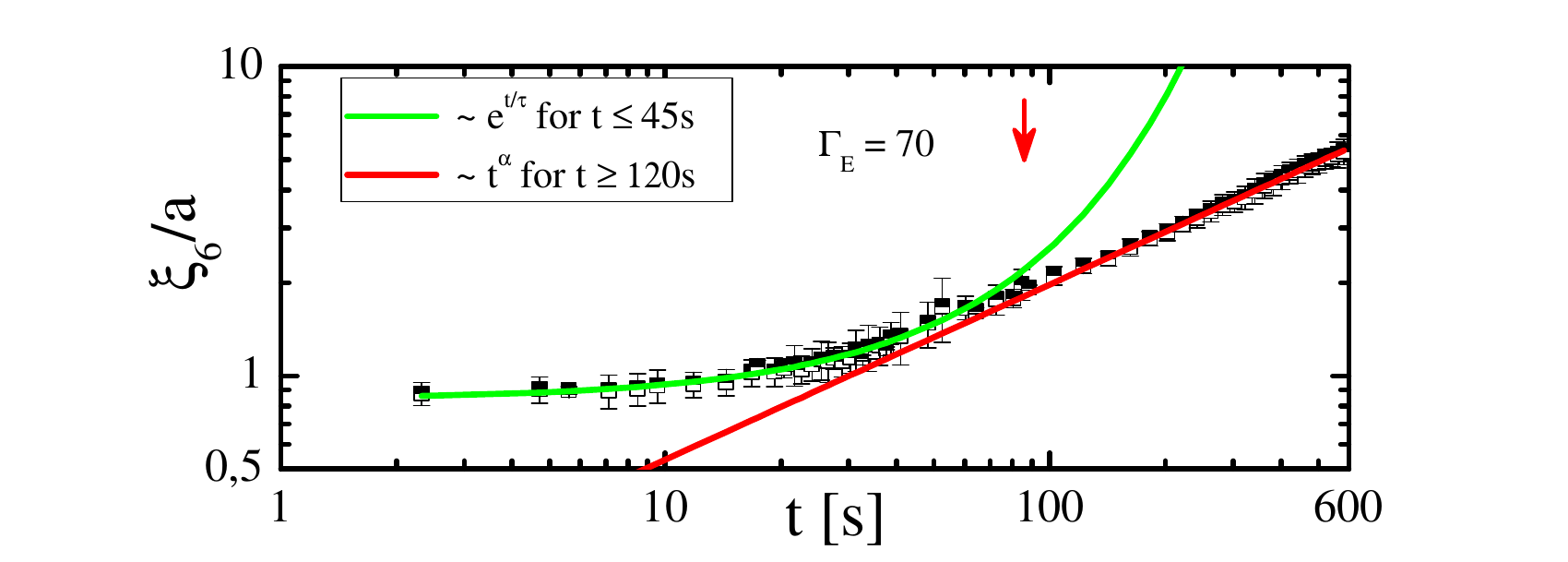}
\includegraphics[width=1.\linewidth]{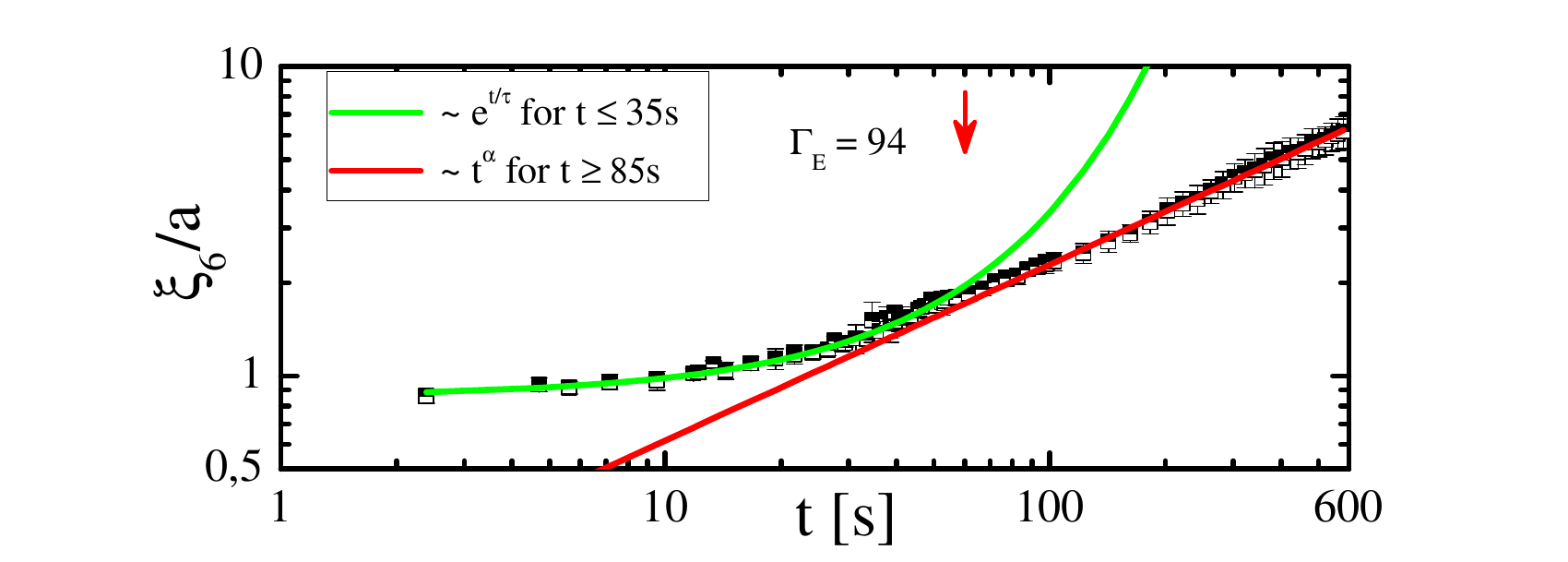}
\includegraphics[width=1.\linewidth]{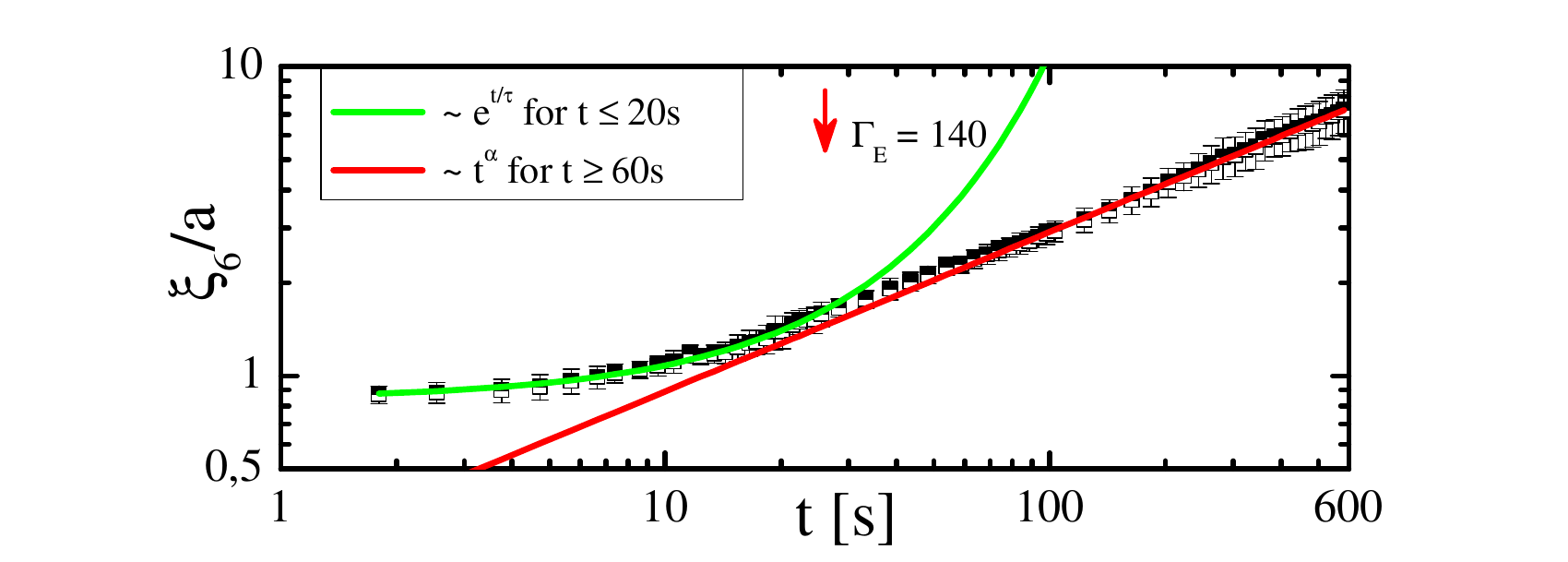}
 \caption{\label{fig02} Orientational correlation length as function of time in a log-log-plot for different quench depth. Shortly after the quench, the growing behavior is exponential and switches to algebraic later. The quench depth increases from top to bottom. Note the decreasing cross-over time (red arrows as guide for the eye) for deeper quenches.
 }
\end{figure}
To analyze the dynamics of locally broken symmetry with respect to Kibble-Zureck mechanism, Figure \ref{fig02} shows the increase of the bond-order correlation length $\xi_6$ for low supercooling ($\Gamma_F = 63$), intermediate ($\Gamma_F = 70 \mathrm{ and } 94$), and deep supercooling ($\Gamma_F = 140$). We use $\xi_6$ as a measure for the average size of symmetry broken regions. In the supplemental material we show, that the standard pair correlation length and the inverse defect density give qualitatively the same results \cite{supp_mat}. In the early state, we observe an exponential grow of the domain size followed by an algebraic one for all magnitudes of investigated supercoolings. The crossover from exponential to algebraic shifts so shorter times for deeper quenches. For the Lambda-transition of $He^4$ Zurek proposed an algebraic decrease of the inverse defect density after the quench \cite{Zurek1993}. For the XY-model a power-law increase of the correlation length with a small logarithmic correction is predicted \cite{Jelic2011} and in dusty plasma an algebraic increase was found \cite{Hartmann2010}. Whether or not the exponential growing of average domain sizes in the early state is a fingerprint of exponentially diverging correlation length in KTHNY-melting can not be answered at present and should be topic of theoretical investigations. In Fig.~\ref{fig02} the cross-over appears for quite small values of the average correlation length (order of unity), even if individual domain sizes e.g. $120~sec$ after a quench to $\Gamma_F = 140$ are already extend over several particles (comparing Fig.~\ref{fig01} and Fig.~\ref{fig02} $\xi_6$ is 'short eyed' and does not measure the size of individual domains). Therefore we introduce a criterium for locally broken symmetries what furthermore allows to follow and label individual domains in time: we define a particle to be part of a symmetry broken domain if the following three conditions are fulfilled for the particle itself and at least one nearest neighbor:
i) The magnitude of the local bond order field $m_{6_k}$ must exceed $0.6$ for both neighboring particles.
ii) The bond length deviation $\Delta |l_{kl}|$ of neighboring particles $k$ and $l$ is less than $10\%$ of the average particle distance $l_a$.
iii) The variation in bond orientation $\Delta\Theta = |\psi_k-\psi_l |$ of neighboring particles $k$ and $l$ must be less than $2.3^\circ$ in real space (less than $14^\circ$ in sixfolded space). An elaborate discussion of criteria to define crystallinity in 2D on a local scale can be found in \cite{Dillmann2013}. Simply connected domains of particles which fulfill all three criteria are merged to a local symmetry broken domain.
\begin{figure*}
\centering
\begin{tabular}{cc}
\includegraphics[width=0.48\linewidth]{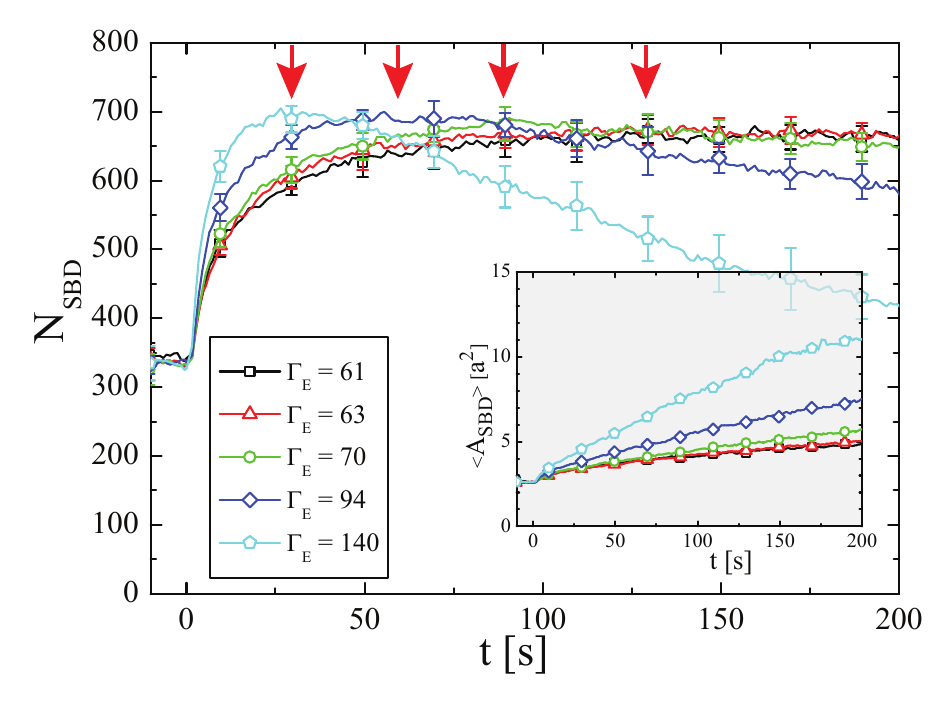} & \includegraphics[width=0.48\linewidth]{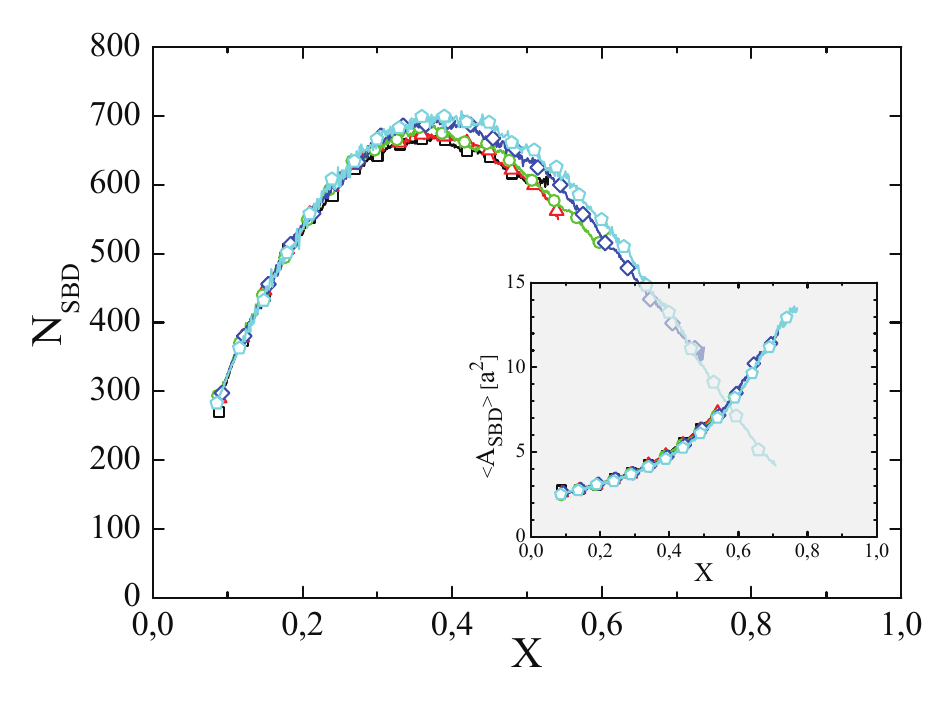} \\
\end{tabular}
\caption{\label{fig03} The left image shows the number of symmetry broken domains (SBD) and the average size of the domains (inset) as function of time for different quench depth (the label is valid for all four figures). The error bars are averages about $10$ independent quenches with about $9000$ particles in the field of view. Whereas the average size of the nuclei grows monotonically as expected, the number of symmetry broken domains first increases to a maximum but then decreases in favor of less but larger domains. The red arrows are a guide to the eye to identify the maxima. The right image shows the same data but plotted as fraction of symmetry broken area (crystallinity, which is implicity a function of time). The curves almost superimpose as function of crystallinity being independent of the quench depth.}
\end{figure*}

Figure~\ref{fig03} [left] shows the average number of local domains as function of time up to $200\;\mbox{sec}$ after the quench and the inset shows the mean size. As expected, the mean size of the nuclei grows monotonically as function of time for all investigated quench depths. The average number of domains (Fig.~\ref{fig03} [left]) first increases but finally decreases in time. As can also been seen in the movies of the supplemental material, fewer but larger domains cover the space as function of time as expected. The maximum sifts to shorter times as function of quench depth - deeper supercooling drives the system faster to the solid state. 
Note that the growing of domains starts immediately after the quench such that no lag time known from classical nucleation theory (CNT) is detectable. (Further discussion in the context of CNT can be found in \cite{supp_mat}.) Figure~\ref{fig03} [right] shows the same data but plotted as function of crystallinity $X$ instead of time. Here, we define crystallinity $X$ (increasing monotonically in time) as the fraction of particles belonging to a symmetry broken domain identified with the three criteria mentioned above. Interestingly all curves almost superimpose as function of crystallinity, re-scaling the time axis to an intrinsic system dependent parameter. Surprisingly, the position and height of average number of symmetry broken domains (in the context nucleation this is called the mosaicity) is independent of the quench depth. It appears when roughly $37\%$ belong to symmetry broken domains. Comparing the maximum on the real time axis (red arrows in Fig.~\ref{fig03} [left]) and the cross-over time in Fig.~\ref{fig02} for different quench depth we propose the following scenario: After a quench, locally symmetry broken domains start to grow exponentially until about $37\%$ of the space is covered. In our 2D system this marks a threshold where domains with different orientation (different phases of the symmetry broken field) start to touch. The following dynamics is dominated by the conversion of the yet untransformed regions of the high symmetry phase marked by an algebraic increase of the bond order correlation length. For finite cooling rates, G. Biroli et al. have argued \cite{Biroli2010} that defect annihilation will alter the classical Kibble-Zurek mechanism leading to two different time regimes after the fall out time, separating critical and classical coarse graining. In the supplemental material we show \cite{supp_mat} that we can resolve the classical coarse graining after an instantaneous quench, too. Large domains grow on the expense of smaller ones with an algebraic increase of the bond-order correlation length with reduced exponent. The latter is due to grain boundary diffusion which was recently described for colloidal mono-layers \cite{Skinner2010,Lavergne2014}.

In the context of Kibble-Zurek-mechanism one would expect all individual grains to grow after the quench which is surprisingly only the case on average. In the supplemental material we show that the shrinking probability $p_s$ of individually labeled domains is always larger than the of growing probability $p_g$. This again rules out a) critical nucleation with a finite nucleation barrier and b) allows for back and forth fluctuations of the broken symmetry regions in 2D systems below the transition temperatures after a quench. Zurek has argued that such fluctuations are suppressed in the vicinity of the transition if the domain size $\Delta A$ times the energy-density differences $\Delta \epsilon$ is less than $k_BT$. In equilibrium this is nothing but another argument for critical slowing down. In 2D systems it is well known that fluctuations play a major role. This is obviously the case not only in equilibrium but also in non-equilibrium situations.

In conclusion, we have investigated the Kibble-Zureck mechanism experimentally for a two-dimensional system with complex order parameter given by the local bond order director field. After a sudden quench we resolve two time regimes at an early state (followed by a third regime due to classical coarse graining, see \cite{supp_mat}). First, the bond order correlation length grows exponentially until roughly $40\%$ of the area belongs to symmetry broken domains with random orientation. While Zurek proposed an algebraic growing for 3D systems, we suggest the exponential behavior being caused by the underlying exponential divergences in KTHNY-melting, being typical for 2D systems. In a second regime an algebraic behavior is observed until no high symmetry phase is left, except grain boundaries separating domains with different phase. The independence of the mosaicity strongly supports the Kibble-Zureck mechanism. We hope our work will stimulate theory and simulations to investigate the phenomena observed in phase transitions far from thermal equilibrium. Further studies in dusty plasma and colloidal mono-layers might shed light on differences and similarities of the Kibble-Zurek mechanism in 2D and 3D systems at sudden quenches or finite cooling rates.

\begin{acknowledgments}
P.K. acknowledges fruitful discussion with S$\acute{\textrm{e}}$bastien Balibar. Financial support from the German Research Foundation (DFG), SFB-TR6 project C2 and KE 1168/8-1 is further acknowledged.
\end{acknowledgments}


\begin{thebibliography}{99}
\bibitem{equilibration_time} The equilibration time to gain a completely flat water/air interface without any gradient of particle numberdensity within the plane is several weeks up to a few months.
\end{thebibliography}

\begin{figure*}
\centering
\includegraphics[width=1\linewidth]{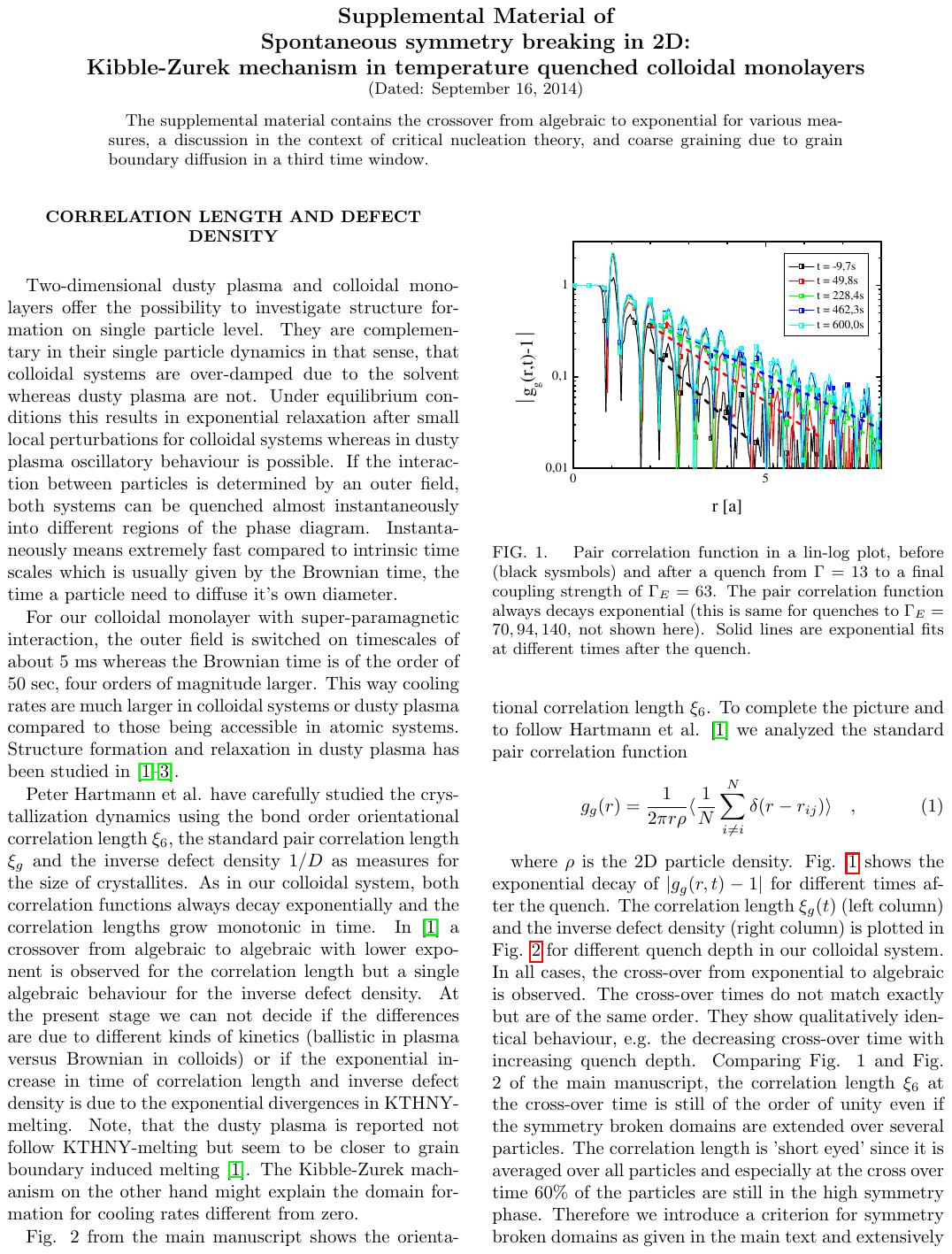}
\end{figure*}
\newpage

\begin{figure*}
\centering
\includegraphics[width=1\linewidth]{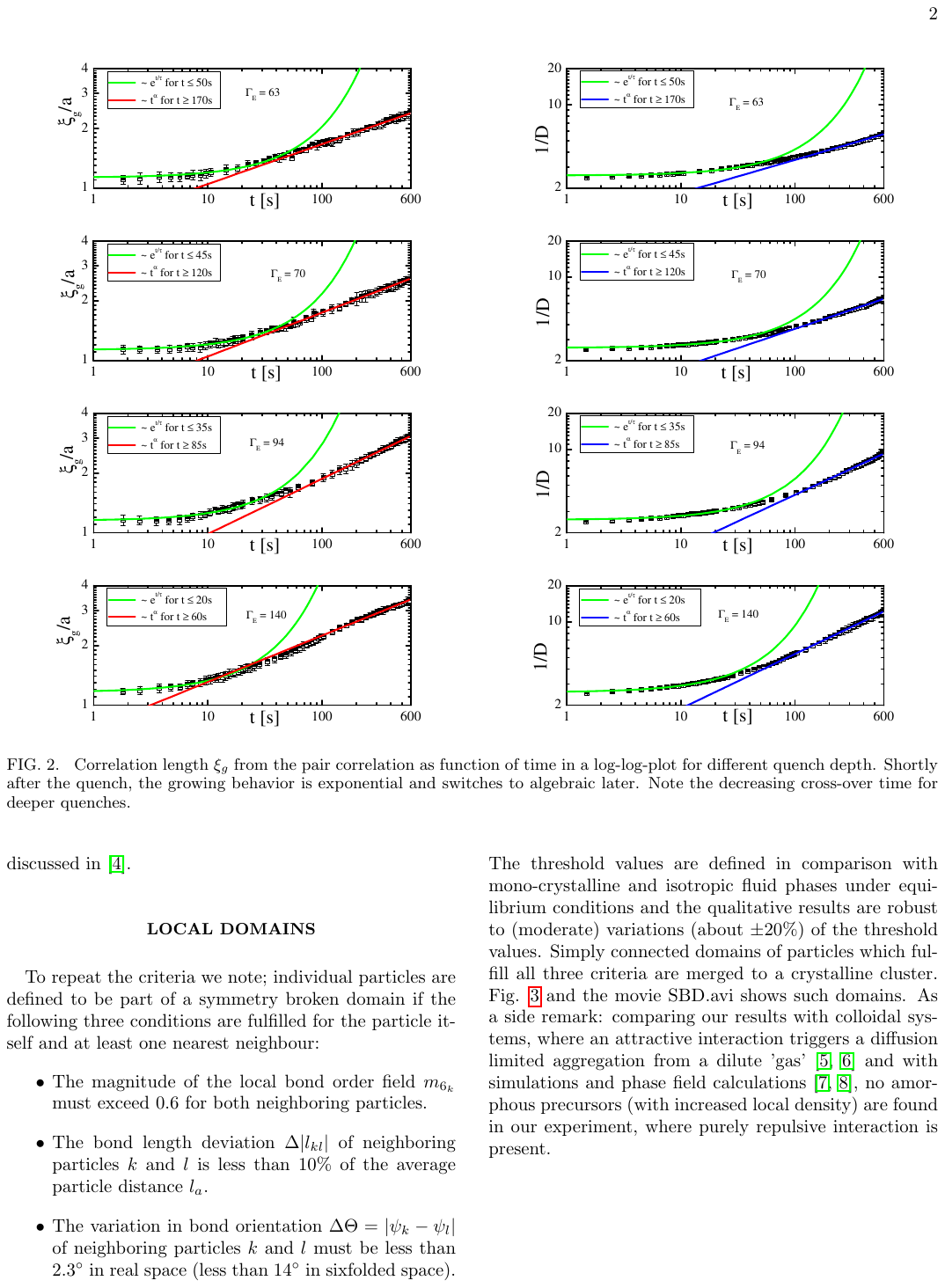}
\end{figure*}
\newpage

\begin{figure*}
\centering
\includegraphics[width=1\linewidth]{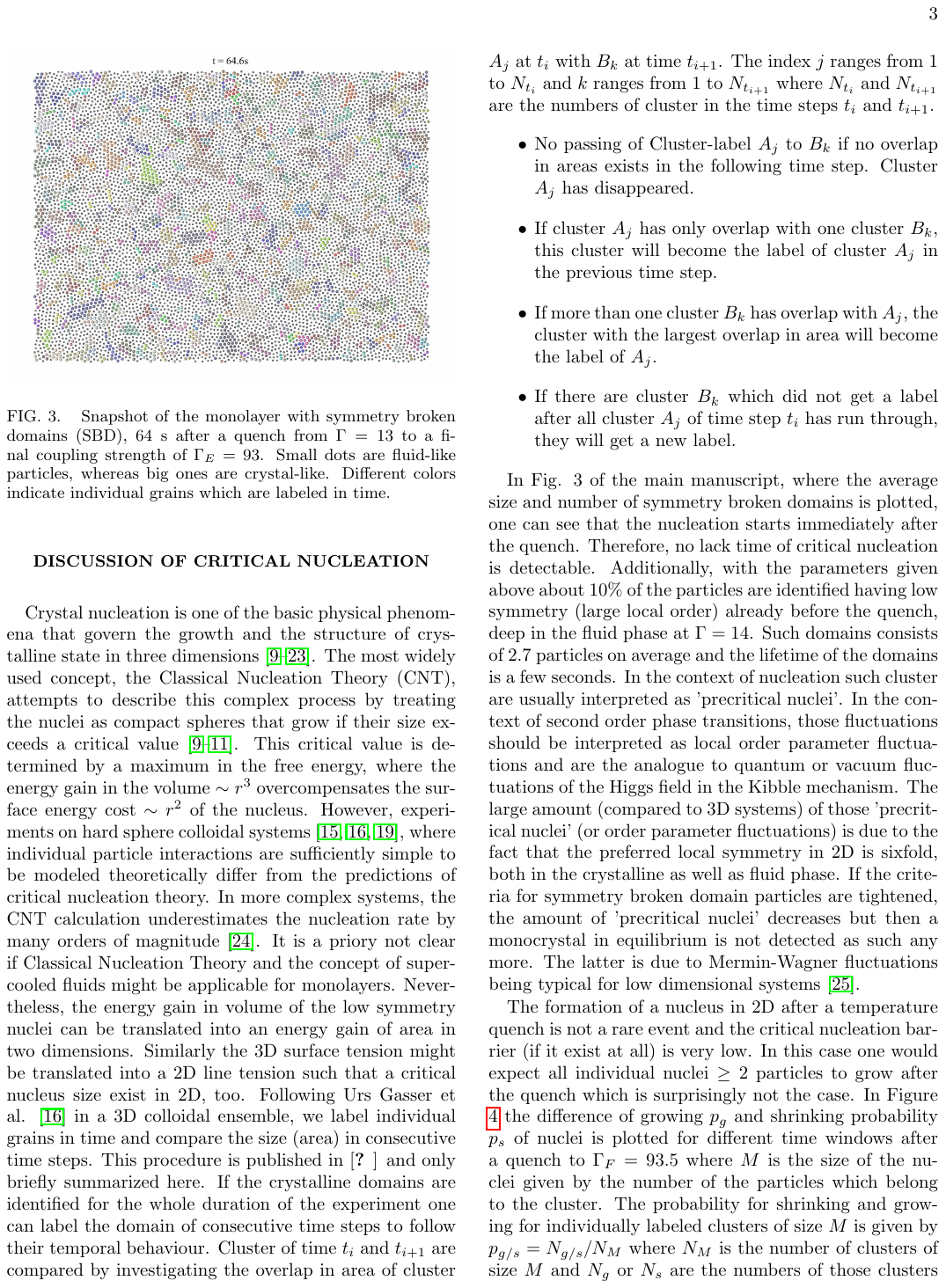}
\end{figure*}
\newpage

\begin{figure*}
\centering
\includegraphics[width=1\linewidth]{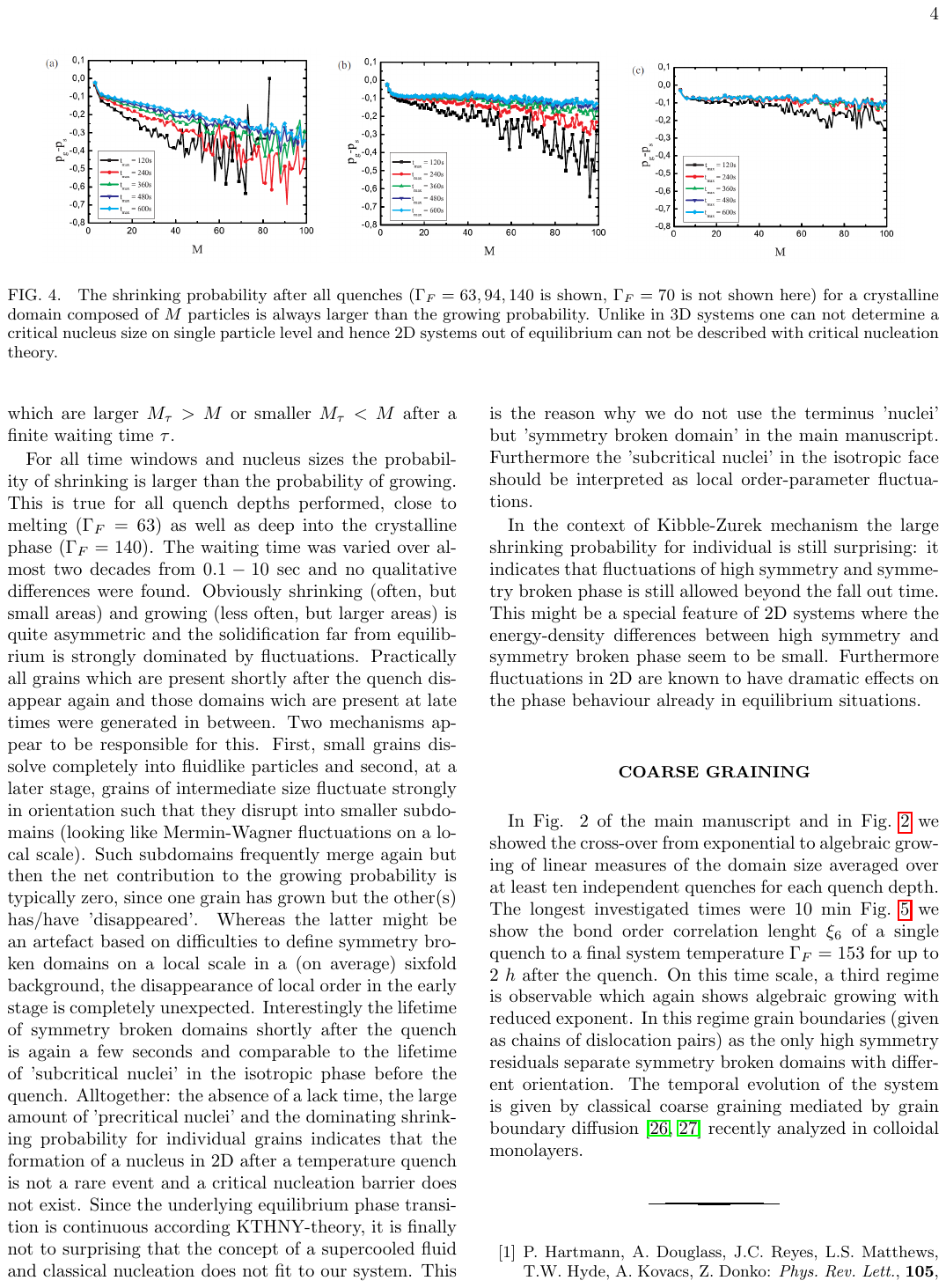}
\end{figure*}
\newpage

\begin{figure*}
\centering
\includegraphics[width=1\linewidth]{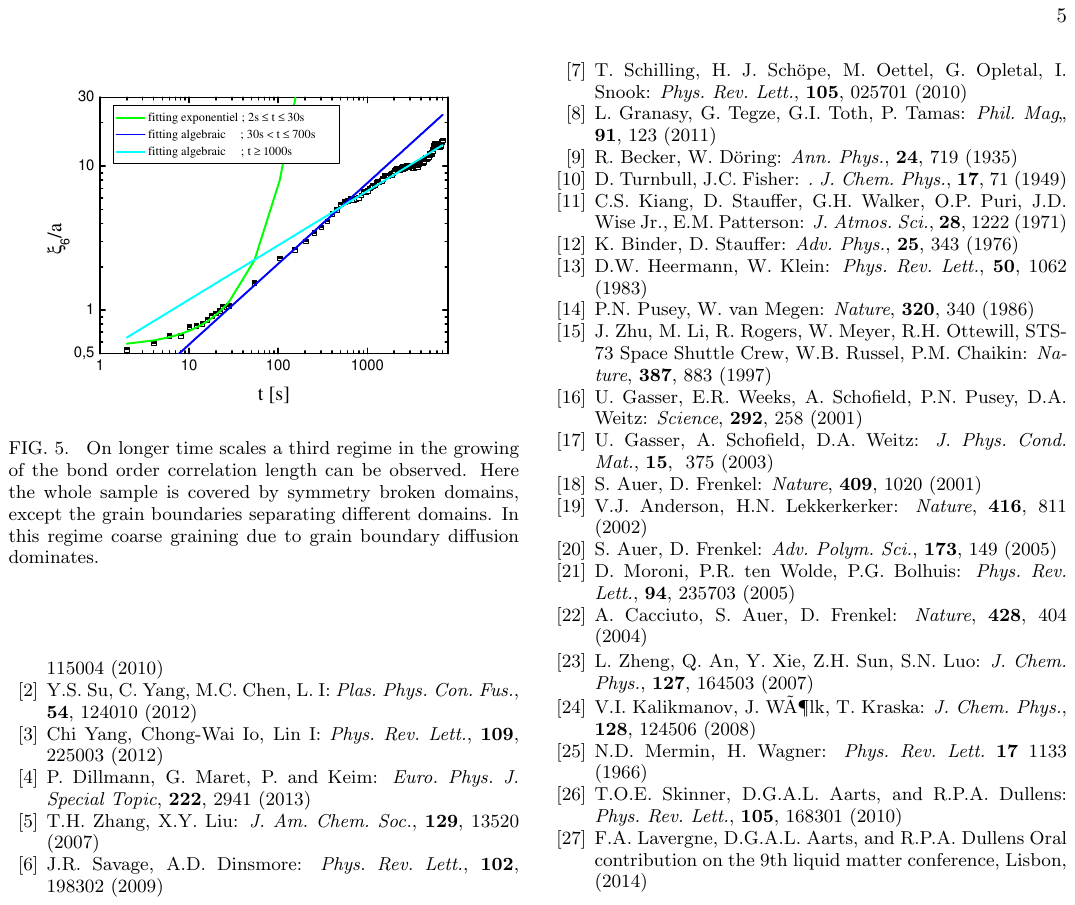}
\end{figure*}

\end{document}